# Assessing the accuracy of individual link with varying block sizes and cut-off values using *MaCSim* approach

Shovanur Haque and Kerrie Mengersen
Queensland University of Technology
shovanur.haque@hdr.qut.edu.au, k.mengersen@qut.edu.au

**Abstract**

Record linkage is the process of matching together records from different data sources that belong to the same entity. Record linkage is increasingly being used by many organizations including statistical, health, government etc. to link administrative, survey, and other files to create a robust file for more comprehensive analysis. Therefore, it becomes necessary to assess the ability of a linking method to achieve high accuracy or compare between methods with respect to accuracy. In this paper, we evaluate the accuracy of individual link using varying block sizes and different cut-off values by utilizing a Markov Chain based Monte Carlo simulation approach (*MaCSim*). *MaCSim* utilizes two linked files to create an agreement matrix. The agreement matrix is simulated to generate re-sampled versions of the agreement matrix. A defined linking method is used in each simulation to link the files and the accuracy of the linking method is assessed. The aim of this paper is to facilitate optimal choice of block size and cut-off value to achieve high accuracy in terms of minimizing average False Discovery Rate and False Negative Rate. The analyses have been performed using a synthetic dataset provided by the Australian Bureau of Statistics (ABS) and indicated promising results.

# 1. Introduction

Record linkage is the process of identifying and combining the same entity that belongs to different data sources when unique identifiers are unavailable (Fellegi and Sunter, 1969). The basic method of record linkage uses information across data files, for example, name, address, age, sex. etc that are typically not considered as unique identifiers of entities, to link the same entities. Record linkage can be used to identify duplicates within or across files to avoid possible erroneous inflated estimates of entities in different categories.

The term record linkage came mainly from the area of public health and particularly from epidemiological and survey applications where it has been used for over a decade. For example, a linked dataset obtained by combining a primary care chronic disease register with hospital inpatient databases, was used for the analysis of determining the prevalence rates (in 2005) of chronic diseases for the remote indigenous population of the Northern Territory of Australia (Zhao et al., 2008). Statistical organisations link administrative, survey and census files in order to create a more comprehensive file in terms of information. For example, Australian Bureau of Statistics linked person records in its 2006 and 2011 Census of Population and Housing to evaluate changes in characteristics of cohorts over time (Zhang and Campbell, 2012).

As massive amounts of data are now being collected by organizations in the private and public sectors, the requirement of linked data from different sources is also increasing. Linking and analysing relevant entities from various sources provide benefits to businesses and government organizations. However, this huge amount of data may also increase the likelihood of incorrectly linked records among databases. Consequently, to make valid inferences using a linked file, it has become increasingly important to have effective and efficient methods for linking data from different sources.

In the case of unavailability of a unique identifier, records are linked probabilistically. The variables used for connecting records are generally called linking variables or linking fields. Each record pair is compared with respect to a set of one or more linking variables (identifying information) and is given a weight based on the likelihood that they are a match. This weight is determined by assessing each linking variable for agreement or disagreement, and then assigned a weight based on this assessment. The assumption of conditional independence of agreement on linking variables within matches and non-matches allows a composite weight to be calculated for each record pair as a simple sum of the individual linking variable weights. Finally, a decision rule based on cut-off values determines whether the record pair is linked, not linked or further considered as a possible link (Fellegi and Sunter, 1969).

As number of possible record pairs depends on the size of the two files, comparing and calculating weights of each pair can cause a significant performance bottleneck for large data files. Instead, record pairs that are similar in terms of basic characteristics are compared within blocks, thus reducing the number of comparisons required and the corresponding computation time.

Increasing the number of blocks improves computational performance since the number of comparisons of record pairs in small groups become smaller, but it also increases the risk of having potential matching records in different blocks, hence decreasing the accuracy. On the other hand, fewer numbers of blocks may result in improved accuracy, but the computational performance will decrease due to the larger number of record pairs to be compared in each block. It is thus critical to find an efficient trade-off between accuracy and performance. This motivates an investigation of the linkage accuracy for varying block sizes in order to achieve an optimal block size.

In probabilistic linking, a matched pair must have a weight which is greater than a specified cut-off value to be declared as a link. It is possible to choose a cut-off value that ensures an acceptably high proportion of correct links (Herzog, Scheuren and Winkler, 2007). The cut-off values should be chosen in such a way as to minimize the number of possible links that cannot be determined at specified levels of errors and require further investigation via clerical review, while balancing the trade-off between the number of false positives and false negatives (Fellegi-Sunter, 1969).

A manual inspection of the weight distribution can help to understand the range of possible weights for matched and non-matched record pairs and to estimate errors. In this paper, we use synthetic data in order to assess the probability of matching and non-matching pair agreement on every linking variable used for linking. The frequencies with which the errors of linking variables occur help us to investigate an optimal cut-off value where we obtain highest accuracy.

The paper is organised as follows. Section 2 describes *MaCSim* approach. *MaCSim* is a Markov chain based Monte Carlo simulation approach that has been developed in a previous work (Haque et al. 2020) and is utilized for the work in this paper. Different quality measurements in the literature are defined in Section 3. Sections 4 and 5 briefly describe the blocking method and cut-off value respectively. In Section 6, we present the analysis of a case study based on synthetic data. Accuracy calculations for different block sizes and varying cut-off values with error estimation are obtained and compared. The paper concludes with a summary in Section 7.

## 2. *MaCSim* approach

This Section gives an overview of the *MaCSim* approach which has been developed in a previous work (Haque et al. 2020).

### 2.1  The purpose of developing *MaCSim*

When there is a task to link two files, it is hard to decide which method to use for linking. Since these are new files, there is no way to measure the accuracy after linking without further review. *MaCSim* can assist in the evaluation of which method will give higher accuracy to link these files. The method can be used as a tool to assess a linking method, or to compare other linking methods. Based on the obtained accuracy results, the user can decide on a preferred method or evaluate whether it is worth linking the two files at all.

### 2.2  *MaCSim*

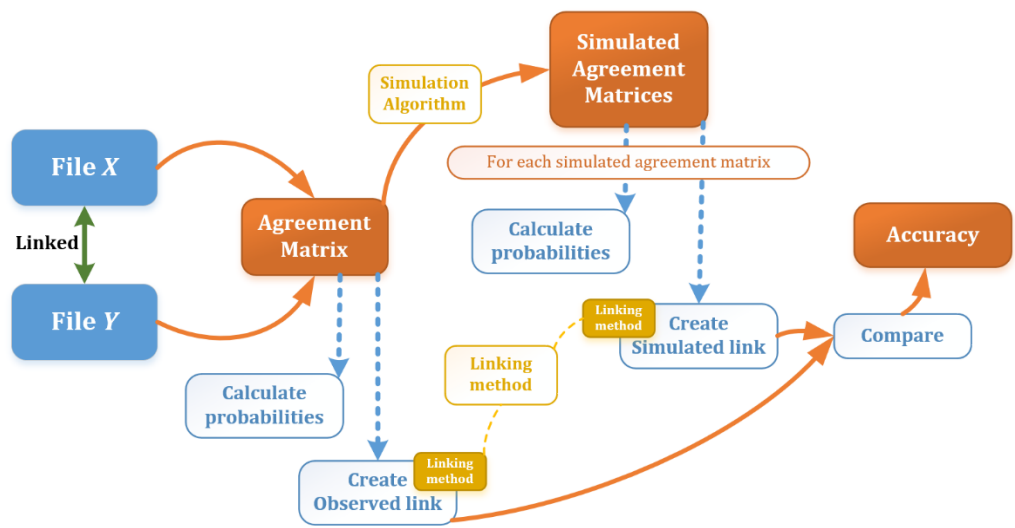

**Figure 1: *MaCSim***

In our previous work, we developed a Markov Chain based Monte Carlo simulation method (*MaCSim*) for assessing a linking method. As we see in the Figure 1, *MaCSim* utilizes two linked files that have been previously linked on similar types of data to create an agreement matrix. From this agreement matrix the necessary parameter values are calculated and links are constructed using a defined linking method. The agreement matrix is then simulated using a defined algorithm developed for generating re-sampled versions of the agreement matrix. In each simulation with the simulated data, records are re-linked using the same linking method. The simulated link is then compared with the observed link and the accuracy of the individual link is calculated. This ultimately provides an evaluation of the accuracy of the linking method that has been followed to link the records.

## 2.3 Creating agreement matrix $A$

An agreement matrix, $A$, is created from the two files to be linked, $X$ and $Y$, where

$$A = (A_{ijl}); \quad i = 1, \ldots R_X, \quad j = 1, \ldots R_Y, \quad l = 1, \ldots, L,$$

is a three-dimensional array denoting the agreement pattern of all linking variables across all records in the two files. Here, $A_{ijl} = 1$ if the $lth$ linking variable value for record $i$ of file $X$ and record $j$ of file $Y$, are the same; $A_{ijl} = -1$ if these values are not the same and $A_{ijl} = 0$ if either or both the values are missing. Therefore, an agreement matrix contains agreement values 1, -1, and 0, which are the comparison outcome between record pairs of the two files to be linked. In the comparison stage, each linking value for a record pair from the two files is compared; the result is a ternary code, 1 (when values agree), -1 (when values disagree) and 0 (when either or both values are missing). According to these codes, each linking variable is given a weight using the probabilities $m$, $u$ and $g$. $m$ is the probability that the variable values agree when the record pair represents the same entity; $u$ is the probability that the variable values agree when the record pair represents two different entities, and $g$ is the probability when the variable values are missing from either or both records in the pair.

For simplicity of notation we assume that $A_{iil}$ represents the agreement value of the $lth$ linking variable for the true matched record pair in both files.

## 2.4 Simulating agreement matrix $A$

In order to assess standard errors for estimates deriving from analysis of the linked data, it is of interest is to generate re-sampled versions of the agreement matrix $A$ in such a way as to preserve the underlying probabilistic linking structure. For this purpose, the *MaCSim* algorithm develops a Markov Chain $\{A^{(n)}\}_{n=0,1,2,\ldots}$ on $A$={set of possible agreement pattern arrays}, with $A^{(0)} = A$, the observed agreement pattern array for the files $X$ and $Y$. The key step is to simulate the observed agreement matrix $A$ to create $A^*$ which includes all the simulated agreement matrices.

## 2.5 *MaCSim* simulation algorithm

The structure of the transition probabilities for the MCMC algorithm employed by *MaCSim* is outlined. Given the current state of the chain, $A^{(n)}$, the next state, $A^{(n+1)}$, is constructed as follows:

*Step 1*: Initially, set $A_{ijl}^{(n+1)} = A_{ijl}^{(n)}$ for all $i, j,$ and $l$.

*Step 2*: Randomly select values of $i \in \{1, \ldots, R_X\}$ and $l \in \{1, \ldots, L\}$.

*Step 3*: If

a) $A_{iil}^{(n)} = 1$, change $A_{iil}^{(n+1)}$ to $-1$ with probability $p_1$.

b) $A_{iil}^{(n)} = -1$, change $A_{iil}^{(n+1)}$ to 1 with probability $p_2$.

*Step 4*: For each $j \neq i$, if

a) $A_{iil}^{(n)} = 1$ & $A_{iil}^{(n+1)} = -1$, then

　　i) If $A_{ijl}^{(n)} = 1$, change $A_{ijl}^{(n+1)}$ to $-1$.

　　ii) If $A_{ijl}^{(n)} = -1$, change $A_{ijl}^{(n+1)}$ to 1 with probability $q_1$.

b) $A_{iil}^{(n)} = -1$ & $A_{iil}^{(n+1)} = 1$ then

　　i) If $A_{ijl}^{(n)} = 1$, change $A_{ijl}^{(n+1)}$ to $-1$.

　　ii) If $A_{ijl}^{(n)} = -1$, change $A_{ijl}^{(n+1)}$ to 1 with probability $q_2$.

c) $A_{iil}^{(n)} = -1$ & $A_{iil}^{(n+1)} = -1$ then

　　If $A_{ijl}^{(n)} = -1$, change $A_{ijl}^{(n+1)}$ to 1 with probability $q_3$.

The values of $p = (p_1, p_2)$ and $q = (q_1, q_2, q_3)$ are determined to ensure the stationary distribution of the chain has the desired structure. Thus, this Markov chain can be used to generate an appropriate set of re-sampled $A$ values.

## 2.6 Maintaining internal agreement consistency

The transition structure is designed to replicate the circumstances whereby a random element of file $X$ is selected and then a change in its value for the $l$th linking variable is made with probability based on its current agreement status with its corresponding partner in the opposite file. If a change does occur, this will have a consequent effect on the agreement patterns in the associated non-matching record pairs. For the matched record pair, if the agreement value of the selected linking variable is changed, that is, from 1 to -1 or -1 to 1 with probability $p_1$ and $p_2$ as in steps $3(a)$ and $3(b)$, then this will have a consequent effect on the agreement patterns in the associated non-matching record pairs. Therefore, if the agreement value of the associated non-matching record pair is 1 then we must change it to -1 as they can no longer agree ($4ai$ & $4bi$). However, if the agreement value of the associated non-matching record pair is -1, then we change it to 1 with probability $q_1$, $q_2$ and $q_3$ because now the value may or may not agree ($4aii, 4bii$ & $4c$). With this underpinning, the internal consistency patterns of agreement will be maintained.

## 2.7 Maintaining marginal distributions

In addition to internal agreement consistency, the chain maintains the required probabilities of agreement for both matched and non-matched records across the two files. This requires appropriate selection of the transition probability parameters $p = (p_1, p_2)$ and $q = (q_1, q_2, q_3)$. In particular, we require the probability that linking variable values for matched record pairs agree remains $m_l$. That is, $Pr\{A_{iil}^{(n+1)} = 1\} = m_l$. Choosing appropriate values for the $q$ parameters arises from the requirement to maintain the probability of agreement between values of the linking variable among non-matched records. In other words, we must ensure that $Pr\{A_{ijl}^{(n+1)} = 1\} = u_l$.

To maintain the marginal probabilities of matching, we choose the transition probability parameters as follows:

$$p_1 = \begin{cases} (1 - m_l - g_l)/m_l & \text{if } u_l \leq 0.5(1 - g_l) \\ (1 - m_l - g_l)(1 - u_l - g_l)/\{m_l(3u_l + g_l - 1)\} & \text{otherwise} \end{cases}$$

$$p_2 = p_1 m_l/(1 - m_l - g_l)$$

$$q_1 = q_2 = \begin{cases} u_l/(1 - u_l - g_l) & \text{if } u_l \leq 0.5(1 - g_l) \\ 1 & \text{otherwise} \end{cases}$$

$$q_3 = 1.$$

The detailed derivation of these values is provided in our previous paper (Haque et al. 2020, submitted).

## 2.8 Calculating $m, u$ and $g$ probabilities

To recap, $m$ is the probability that the variable values agree when the record pair represents the same entity; $u$ is the probability that the variable values agree when the record pair represents two different entities, and $g$ is the probability when the variable values are missing from either or both records in the pair.

For each linking variable $l$, $m_l$, $u_l$ and $g_l$ are calculated in the following way:

$\widehat{m_l}$ = number of values that agree for matched record pairs/total number of matched record pairs.

$\widehat{u_l}$ = number of values that agree for non-matched record pairs/total number of non-matched record pairs.

$\widehat{g_l}$ = total number of record pairs of which one or both values are missing/total number of possible record pairs.

## 2.9 Creating observed link

After comparing each linking variable value for a record pair from the two files, each linking variable is given a weight according to the agreement values in the agreement matrix $A$, using the probabilities $m_l, u_l,$ and $g_l$. For any $(i,j)$-th record pair and any linking variable $l$, if the agreement value is 1 (i.e. $A_{ijl}=1$) then the weight is calculated using $w_{ijl} = log\left(\frac{m_l}{u_l}\right)$; if the value is -1 (i.e. $A_{ijl}=-1$), the weight is calculated using $w_{ijl} = log(1 - m_l - g_l)/(1 - u_l - g_l)$ and for a missing value (i.e. $A_{ijl}=0$), the weight formula is $w_{ijl} = log\,(g_l/g_l) = log\,(1)$.

Since we assume that missingness occurs at random, and therefore has the same chance of occurring in a true matched pair as in a non-match, missing values will not contribute to the weight.

After calculating the weight for each record pair that agree or disagree on a linking variable value explained above, a composite or overall weight, $W_{ij}$ is calculated for each record pair $(i,j)$ by summing individual weights, $w_{ijl}$ over all linking variables $l$ for that pair using the following formula:

$$W_{ij} = \sum_l w_{ijl}$$

Once weights of all record pairs, $W_{ij}$ are calculated, the observed links are created following the steps of defined linking method below:

a. First, all record pairs are sorted by their weight, from largest to smallest.
b. The first record pair in the ordered list is linked if it has a weight greater than the chosen cut-off value.
c. In all the other record pairs that contain either of the records from the associated record pair that have been linked in step b, are removed from the list. Thus, possible duplicate links are discarded.
d. Go to step b for the second record and so on until no more records can be linked.

To calculate the accuracy of the linking method above, these observed links are compared with simulated links obtained by using the same linking method (Haque et al. 2020, submitted). Note that the maximum number of records that are considered as links will be less than or equal to the number of records in the smaller file since it contains maximum number of possible matches. With the varying cut-off values we can accept record pairs that have weights greater than the corresponding cut-offs. However, for any cut-off, the number of declared links should not exceed the total number of true matches since duplicate links are not allowed.

## 3. Estimating accuracy

Measuring the quality of the linked file is important. There are many quality measures that can be defined based on a concordance matrix (see Table 1),

**Table 1 Concordance table for binary classifications**

|  |  | Link status | | |
|---|---|---|---|---|
|  |  | Link | Non-link |  |
| **True match status** | **Matches** | True positive (TP) | False negative (FN) | Total matches (TM) |
|  | **Non-matches** | False positive (FP) | True negative (TN) | Total non-matches (T$\overline{\text{M}}$) |
|  |  | Total links (TL) | Total non-links (T$\overline{\text{L}}$) | Total record pairs (N) |

Sensitivity and specificity are common statistical measures of the performance of a classification test. In our setting, sensitivity measures the proportion of links that are correctly identified as true matches by the estimate as well. From Table 1, sensitivity is calculated by

$$TP*100/(TP+FN)$$

Specificity measures the proportion of non-links that are correctly identified as non-matches by the estimate as well. From Table 1, specificity is calculated by

$$TN*100/(TN+FP)$$

Several methods for selecting a true match and for estimating linkage errors are available in the literature (Bartlett, Krewski, Wang and Zielinski, 1993; Armstrong and Mayda, 1993; Belin, 1993; Belin and Rubin, 1995 and Winkler, 1992 & 1995). The methods in the literature for estimating linkage error rates, consider different error components in the estimation and different approaches for linkage. The range of alternatives proposed in these articles indicates that comparing estimates obtained from different methods is complicated.

According to the approach by Fellegi-Sunter (1969), an estimate of false positive linkage error rate is obtained by

$$\hat{\mu} = Pr(A_1|U).$$

Here, $A_1$ represents links when $U$ is a non-match set. This is estimated by the ratio of false links to total non-matches. Similarly, an estimate of false negative linkage error rate is

$$\hat{\lambda} = Pr(A_3|M),$$

where $A_3$ represents non-links when $M$ is a match set. This would be estimated by the ratio of missed links to total matches. Estimation of these quantities ($\hat{\mu}, \hat{\lambda}$) depends on match and non-match probabilities for each linking variable. These probabilities are assumed to be independent of the actual value of the linking variable.

Belin and Rubin (1995) noticed that Fellegi-Sunter's approach does not capture actual error rate since this conditional independence assumption does not usually hold. They suggest calculating a 'false match rate' which is estimated by,

$$1 - (number\ of\ true\ links/total\ links).$$

In their model, the distribution of observed weights is considered to be a mixture of two distributions, one for matched pairs and the other for non-matched pairs. These distributions are estimated by fitting transformed normal curves to the record pair weights. The method is robust to independence assumptions. Although their method estimates asymptotic standard errors, they rely on clerical review for parameter estimation. Another drawback of the method is that the distributional assumptions may not be valid when the weight distribution for either the matched or non-matched set is multimodal. Since the method uses training data for parameter estimation, without a good training data set, these input parameter estimates for the mixture model may be poor which will affect the estimated error rates (Winglee, Valliant and Scheuren, 2005). In a later paper, Winkler (2006) noted that this method performs well when the curves are well separated for matches and non-matches.

Winglee, Valliant and Scheuren (2005) designed a simulation approach, *Simrate*, for estimating $\mu$ and $\lambda$. Their method uses the observed distribution of data in matched and non-matched pairs to generate a large simulated set of record pairs. They assign a match weight to each record pair following specified match rules, and use the weight distribution for error estimation. The simulated distribution is used to select an appropriate cut-off for estimating the error rates $\mu$ and $\lambda$.

For the measurement of linkage quality, Christen (2012) suggests precision, defined as the proportion of links that are matches. In this paper, for a quality measure, our primary interest is to calculate the accuracy of an individual link. We calculate the correct link proportion among the true matches and expressed it as accuracy. The challenge of measuring linkage accuracy in practice is to determine the true match status for each record pair. However, for the synthetic data used in our case study we know the true matches (TM) and we can calculate the accuracy of link for each record in every simulation.

## 4. Blocking

The variables that are used for blocking need to be chosen carefully. An erroneous value in blocking variable can insert a component record of a true match into different blocks. As records that do not agree on the blocking variable will automatically be classified as non-matched, this means that with such blocking, true matches will never be able to be correctly linked. Blocking variable should have few missing values. The values of the blocking variable should ideally be uniformly distributed as the most frequent values determine the size of the largest blocks. For example, 'age' is generally not uniformly distributed and doesn't make a good blocking scheme alone (Jaro, 1995). However, when 'age' is combined with other variables and used together as a blocking variable, a better result may be achieved. Increasing the number of blocks gives us higher performance regarding computation, but it also increases the risk of having potential matching records in different blocks, consequently decreasing the accuracy. On the other hand, fewer numbers of blocks may give higher accuracy, but the computational performance will decrease due to the higher number of records in each block. Therefore, it is crucial to find an efficient trade-off between accuracy and performance. In this paper, we investigated the average accuracy of each record with different block sizes.

## 5. Cut-off values

In most published studies, only one cut-off value is chosen and the reason behind the chosen cut-off value varies from study to study. For example, it can be chosen based on manual inspection of the weight distribution (Tromp et al., 2009; Lyons et al., 2009) or based on past experiences (Gorelick et al., 2007). Belin (1993) demonstrated that the choice of cut-off value is crucial. In the classification table (Table 1) for matches and links, total match is TM=(TP+FN) while the total links is TL=(TP+FP). By definition, false positive rate is FPR=FP/(FP+TN) and false negative rate is FNR=FN/(TP+FN). In general, the TN (true negative) cell contains a larger number of records than any other cell in the table, which implies a very small FPR. The false discovery rate FDR is defined as FDR=FP/(TP+FP) i.e., the proportion of declared links that are actually non-matches. In this paper, we investigated average accuracy and also inspected average False Discovery Rate (FDR) and average False Negative Rate (FNR) for a range of cut-off values. It is desirable to have a cut-off value where we could obtain high accuracy and also have low FNR and FDR at the same time.

## 6. Analysis

### 6.1 Data

A synthetic dataset based on realistic data settings received from the Australian Bureau of Statistics is used for demonstration and analysis to avoid privacy issues associated with using real personal information. A large file $Y$ is generated that comprises 400,000 randomly ordered records corresponding to 400,000 hypothetical individuals. Then, from file $Y$, the first 50,000 records are taken to form file $X$. Every record has eight data fields. These fields are: RECID (Record Identifier), SA1 (Statistical Area 1), MB (Meshblock), BDAY (Birth day), BYEAR (Birth year), SEX (Male/Female), EYE (Eye colour) and COB (Country of birth).

For a record, the value of each variable is generated independently except the value of Country of Birth (COB). For this variable, 300,000 records are assigned a value '1101' for 'Born in Australia' and the remaining 100,000 records are randomly assigned one of about 300 country codes according to the corresponding proportion of people in the 2006 Australian Census. In file $X$, the RECID (Record Identifier) remained matched to the Y file for each record. This makes it easy to identify true matches and non-matches in the linking process. Some values in file $X$ are changed intentionally to simulate errors in linking variables. The value of a variable in file $X$ is changed by replacing it either with a randomly chosen value from the records in file $Y$ or setting the value to 'missing' (Haque et al, submitted).

**6.2 Analysis setup**

In a previous paper (Haque et. al., 2020) we proposed a Markov Chain based Monte Carlo simulation (*MaCSim*) approach to assess a linking method. *MaCSim* utilizes two linked files with known true match status to create an agreement matrix. The agreement matrix is created by comparing all records in the two files using all linking variables. From this initial agreement matrix, the observed links are obtained by applying a specified linking method. Then the agreement matrix is simulated many times using a defined algorithm developed for generating re-sampled versions of the agreement matrix. Utilizing each simulated matrix, we link the files $X$ and $Y$ each time with the same linking method and calculate the accuracy of the link by the correct re-link proportions based on the links before and after simulations. Test results showed robust performance of the proposed method of assessment of accuracy of the linkages. In this present paper, interest focuses on the impact of block sizes on linkage accuracy in order to determine the optimal block size. To address this, we applied our *MaCSim* approach and calculated the accuracy of individual link for different block sizes.

In the linking method employed by *MaCSim*, we calculate the weights for all record pairs and considered pairs as a link if they have weights larger than the cut-off value. All the record pairs that have weights smaller than the cut-off value are considered as non-link. Hence, it is important to set up the cut-off value in such a way that the number of false positives and false negatives are minimized to achieve an efficient trade-off between these two measures. In this paper, we calculated average accuracy and errors for different cut-off values in the *MaCSim* approach.

### 6.3 Average proportion of correct links for varying block sizes

In this analysis, the aim is to evaluate how the average correct link proportion changes with the increasing block sizes. When we block with SA1, we have 1000 blocks in both files. The first block in file *X* has 59 records to be matched with 400 records in file *Y*. We divide these 59 records into 10 sub-blocks: 1-6, 1-12, 1-18, …., 1-59. For each sub-block, we have 400 records in file *Y*. We start with a small block size of 6 records and gradually increase the size of the block with 59 records. For each block, we calculate the average accuracy for each record in 1000 simulations. The result is plotted in Figure 1.

From Figure 2, we can see that average accuracy of a link increases rapidly with increasing block sizes, then asymptotes. The highest accuracy we obtain is 99.5% with block size 24. After that, accuracy stays stable at around 98%. Therefore, block size 24 can be taken as optimal block size where an efficient trade-off between accuracy and performance is achieved.

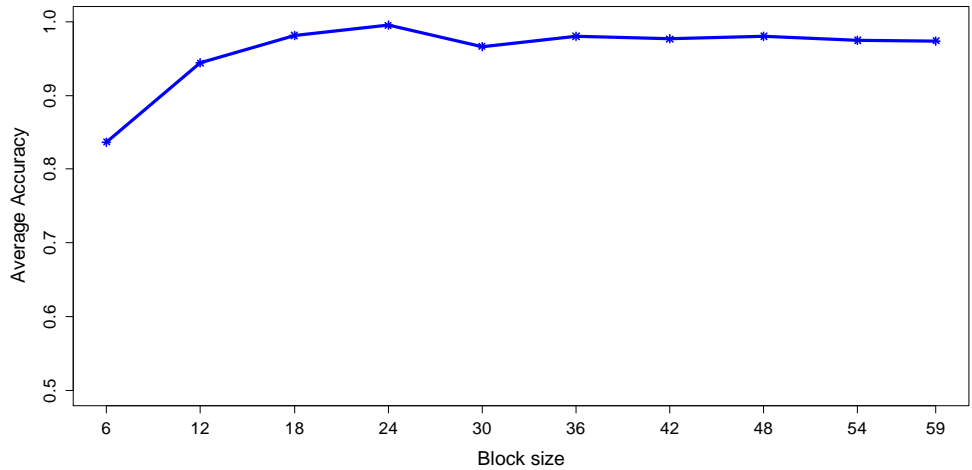

**Figure 2: The average accuracy with changing block sizes**

### 6.4 Average accuracy of correct links with varying cut-offs

In this analysis, we observe the changes in average accuracy of correct links for a range of cut-off values. We considered cut-off values ranges from -15 to +20. For each cut-off value, we calculated accuracy in 1000 simulations and averaged the results. The average accuracy for every cut-off value is plotted in Figure 3.

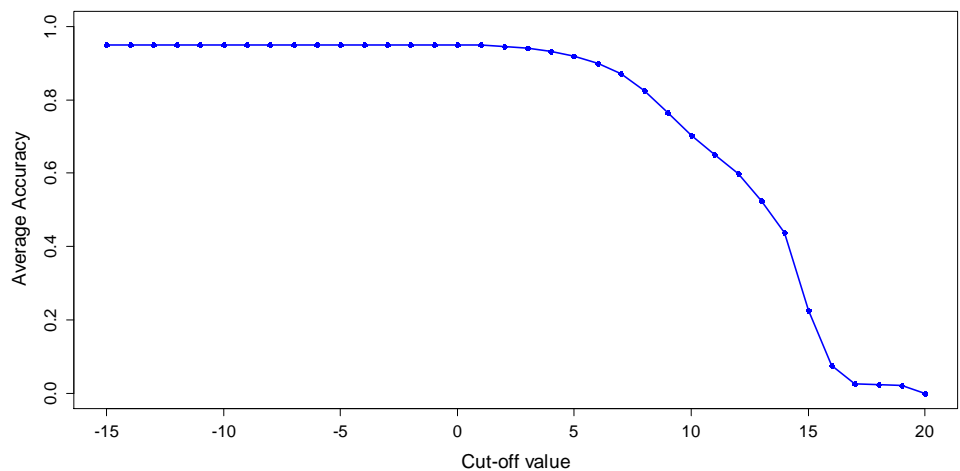

**Figure 3: The average accuracy with changing cut-off values**

From Figure 3, it is observed that for cut-off values up to and including 4, we obtained an accuracy of around 95% for all records in each simulation. From cut-off value 5, average accuracy decreases and ends up at accuracy 0 with the cut-off value 20, which means there is no record pair with weights more than 20.

It is observed that with the smaller cut-off values, the accuracy stays unchanged and high. This is because when we decrease the cut-off value, the number of record pairs with larger weights than the cut-offs increases; which in turn increases the number of true positives (TP). In our calculation, TP/(TP+FN), the denominator (total number of true matches, TM) is known and fixed. Thus, when we decrease the cut-off value, the accuracy increases, and at some point, it reaches the maximum and stays there.

## 6.5 Average False Discovery Rate and False Negative Rate with varying cut-offs

In this section, the average accuracy, FDR and FNR are observed for varying cut-off values. The aim is to find a cut-off at which an efficient trade-off between FDR and FNR will be achieved with high accuracy.

Figure 4 shows the average FDR for each cut-off value in the range from -15 to +20 based on 1000 simulations. The average FDR for every cut-off value is plotted in Figure 4.

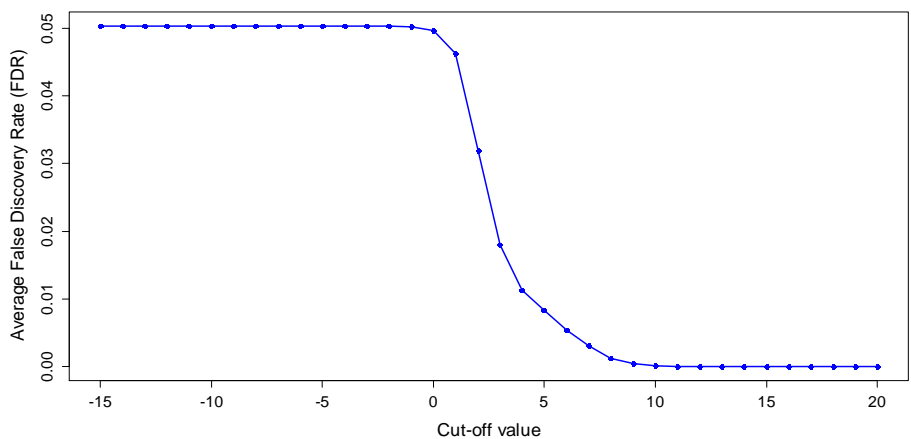

**Figure 4: Average False Discovery Rate (FDR) with changing cut-off values**

From Figure 4, it is observed that the FDR for cut-off values up to and including 1 is 0.05. FDR values decrease from a cut-off value of 2. From a cut-off of 15, the FDR became 0 i.e., there are no estimated links. The largest FDR we obtained is 0.05, meaning that 5% of declared links are truly a non-match. The small FDR value represents high performance of the method. In our previous analyses for assessing the accuracy of individual links, we used a cut- off value of 4. From Figure 4, the FDR for a cut-off 4 is 0.018 or 1.8%.

Figure 5 shows the average false negative rate FNR=FN/(TP+FN) for each cut-off value in the range from -15 to +20 based on 1000 simulations. From this figure, we observe that the average false negative rate (FNR) for cut-off values from -15 to 3 is 0.05. From a cut-off value of 4, the FNR started increasing rapidly to 0.98 for cut-off values of 17 and above. The formula of FNR (=FN/(TP+FN)) implies that TP dominates the value of FNR. When TP increases FNR decreases and when TP decreases, FNR increases. Again, TP decreases with the increasing cut-off value. This is why we obtain large FNR when cut-off value increases.

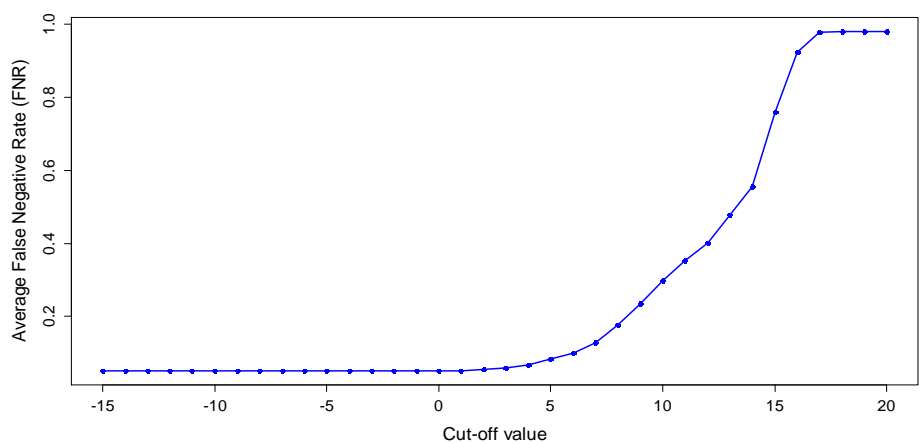

Figure 5: Average False Negative Rate (FNR) with changing cut-off values

From the two analyses (Figure 4 & Figure 5), we obtained FDR values between 0.001 and 0.05, and FNR values between 0.1 and 0.98, for cut-off values from -15 to 20. It is noticeable that between cut-off values 2 and 8, we obtained low estimates for both the FDR and FNR. Again, in the analysis of accuracy for different cut-offs (Figure 3), we achieved higher accuracy when the cut-off value is below 5. In Figure 6, we show the average accuracy, false discovery rate (FDR) and false negative rate (FNR) for varying cut-off values in a single plot.

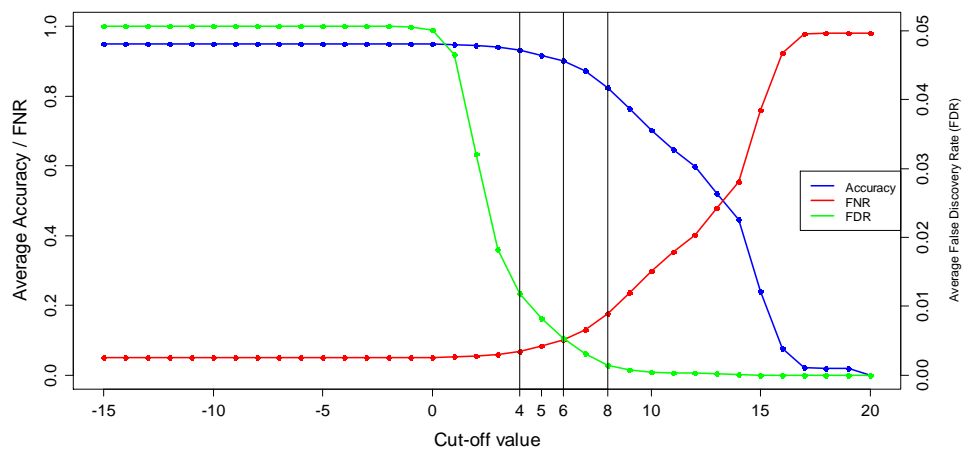

**Figure 6: Average Accuracy, FNR and FDR with changing cut-off values**

When the cut-off value is 6, we have an efficient trade-off between FDR and FNR. However, this is at the expense of accuracy. Indeed, the best trade-off among these three measurements appears to be achieved at a cut-off value of 4 where the accuracy is high and both FNR & FDR are reasonably low.

## 7. Conclusion

In our previous paper, we described a Markov Chain based Monte Carlo simulation approach, *MaCSim*, for assessing linkage accuracy. In the linking process, different size of blocks and different cut-off values plays a vital role in accuracy. It is critical to find optimal block sizes and cut-off values. In this paper, we observed average accuracy by varying block sizes and varying cut-off values. We described the implication of block sizes on performance of linking accuracy. We also described the process of finding an optimal cut-off value by obtaining an efficient trade-off between error estimates and accuracy. The analyses showed good results for finding the optimal cut-off value as well as helping to make decisions on efficient block sizes.

The weight of record pair depends on the number of variables and their agreement or disagreement on variable values. Using our approach for a different dataset with a different set of variables, the optimal cut-off value as well as the efficient block size can be different. The results described in this paper suggest an optimal cut-off and efficient block size on a synthetic dataset of hypothetical individuals. The same approach can be applied to other datasets and will help to find an optimal cut-off value and block size for those datasets.

**References**

Armstrong, J. B. and Mayda, J. E. (1993). Model-Based Estimation of Record Linkage Error Rates. Survey Methodology, 19, 137–147.

Bartlett, S., Krewski, D., Wang, Y. and Zielinski, J.M. (1993). Evaluation of error rates in large scale computerized record linkage studies. Survey Methodology, 19, 3-12.

Belin, T. R. (1993). Evaluation of Sources of Variation in Record Linkage through a Factorial Experiment. Survey Methodology, 19, 13–29.

Belin, T. R. and Rubin, D. B. (1995). A Method for Calibrating False-Match Rates in Record Linkage. Journal of the American Statistical Association, 90, 694–707.

Christen, P. (2012). Data Matching. New York: Springer.

Fellegi, I.P. and Sunter, A.B. (1969). A Theory for Record Linkage. Journal of the American Statistical Association, 64, 1183–1210.

Gorelick, M.H., Knight, S., Alessandrini, E.A., Stanley, R.M., Chamberlain, J.M., Kuppermann, N., Alpern, E.R. (2007). Lack of agreement in pediatric emergency department discharge diagnoses from clinical and administrative data sources. Academic Emergency Medicine 14:646–652.

Haque, Shovanur, Mengersen, Kerrie, Stern, Steven (2020, submitted). Assessing the accuracy of record linkages with Markov chain based Monte Carlo simulation approach. Submitted for publication.

Herzog, T.N., Scheuren, F.J. and Winkler, W.E. (2007). Data Quality and Record Linkage Techniques. Springer: New York.

Jaro, M.A. (1995). Probabilistic Linkage of Large Public Health Data Files. Statistics in Medicine, 14, 91–498.

Lyons, R.A., Jones, K.H., John, G., Brooks, C.J., Verplancke, J.P., Ford, D.V., Brown, G., Leake, K. (2009). The SAIL databank: linking multiple health and social care datasets. Bmc Medical Informatics and Decision Making 9; 3.

Tromp, M., Van, Eijsden M., Ravelli, A., Bonsel, G. (2009). Anonymous non-response analysis in the ABCD cohort study enabled by probabilistic record linkage. Paediatric and Perinatal Epidemiology 23:264–272.

Winkler, W.E. (1992). Comparative Analysis of Record Linkage Decision Rules. In Proceedings of Survey Research Methods Section, American Statistical Association, pp. 829–834.

Winkler, W.E. (1995). Matching and Record Linkage. In Business Survey Methods, eds. B. G. Cox, D. A. Binder, B. N. Chinnappa, A. Christianson, M. J. Colledge, and P. S. Kott, New York: Wiley, pp. 355–384.

Zhao, Y., C. Connors, J. Wright and S. Guthridge (2008). Estimating chronic disease prevalence among the remote aboriginal population of the northern territory using multiple data sources. Australian and New Zealand Journal of Public Health 32, 307–313.

Zhang, G. and Campbell, P. (2012). *Data Survey:* Developing the Statistical Longitudinal Census Dataset and Identifying Its Potential Uses. Australian Economic Review, 45, pp125-133.

Winkler, W.E. (2006). Overview of Record Linkage and Current Research Directions. Statistical Research Division U.S. Census Bureau Washington, DC 20233. Statistical Research Division Report.

Winglee, M., Valliant, R. and Scheuren, F. (2005). A Case Study in Record Linkage*.* Survey Methodology. Vol. 31, No. 1, pp. 3-11.